\documentstyle[preprint,prd,aps]{revtex}
\begin{document}
\title{Multiquark States in a Goldstone Boson Exchange
Model}
\author{Fl. Stancu\thanks{{\it E-mail address:}
fstancu@ulg.ac.be}}
\address{Institute of Physics, B.5,
University of Liege, Sart Tilman, B-4000 Liege 1,
Belgium}

\sloppy

\maketitle
\begin{abstract}
We discuss the stability of multiquark systems containing heavy
flavours. We show that the Goldstone boson exchange model
gives results at variance with one-gluon-exchange models.
\end{abstract}

\section{Introduction}

The study of exotic hadrons formed of more than three quarks and/or
antiquarks ($ q^m \overline{q}^n$ with $ m+n > 3 $) is a natural
development of QCD inspired models. Both theoretical and experimental
interest has been raised so far by particles described by the colour
state ${\left[{222}\right]}_C$. These are the tetraquarks
$q^2\overline{q}^2$ \cite{JA77}, the pentaquarks $q^4\overline{q}$
\cite{GI87,LI87} and the hexaquarks $q^6$ \cite{JA77}.
From theoretical general
arguments \cite{MA93,RI94}, one expects an increase in stability of
multiquark systems if they contain heavy flavours $Q = c$ or $b$.

In the heavy sector, experiments are being
planned at CERN and Fermilab to search for new heavy hadrons and in
particular for doubly charmed tetraquarks \cite{MO96}. Recently ,
the first search for pentaquarks with the flavour content
$uuds\overline{c}$ and $udds\overline{c}$ has just been
reported \cite{AI98}. Within the confidence level of the analyzed
experiments, no convincing evidence for the production of
the above strange pentaquarks has been observed so far.

The theoretical predictions are model dependent. Here we are
concerned with constituent quark models which simulate the
low-energy limit of QCD. We compare results from models where
the spin-dependent term of the quark-quark interaction is described by
the chromomagnetic part of the one gluon exchange (OGE) interaction
\cite{RU75}
with results we obtained from the Goldstone boson exchange
(GBE) model \cite{GL96,GP96,GL97,WA98}.
In this model the hyperfine splitting in hadrons is
due to the short-range part of the Goldstone boson exchange
interaction between quarks, instead of the OGE interaction of
conventional models. The GBE interaction is flavour-dependent
and its main merit is that it reproduces the correct ordering
of positive and negative parity states in all parts of the
considered spectrum. Moreover, the GBE interaction induces a
strong short-range repulsion in the $\Lambda$-$\Lambda$
system, which suggests that a deeply bound H-baryon should
not exist \cite{ST98}. This is in agreement with the
high-sensitivity experiments at Brookhaven \cite{ST97}
where no evidence for H production has been found.

In the stability problem we are interested in the quantity
\begin{equation}
\Delta E =\ E(q^m\overline{q}^n) - \ E_T
\end{equation}
where $E(q^m\overline{q}^n)$ represents the multiquark energy and
$E_T$ is the lowest threshold energy
for dissociation
into two hadrons: two mesons for tetraquarks, a baryon + a meson
for pentaquarks and two baryons for hexaquarks. A negative $\Delta E$
suggests the possibility of a stable compact mutiquark system.

According to Ref. \cite{GL96} there is no meson exchange interaction
between quarks and antiquarks. It is assumed that the
$q\overline{q}$ pseudoscalar pairs
are automatically included in the GBE interaction.
Therefore the light
quark and the heavy antiquark interact via the confinement
potential only
and the model Hamiltonian contains GBE interactions only
between light quarks.

\section{The Hamiltonian}
\label{secstyle}
The GBE Hamiltonian considered below has the form \cite{GP96} :
\begin{equation}
H= \sum_i m_i + \sum_i \frac{\vec{p}_{i}^{\,2}}{2m_i} - \frac {(\sum_i
\vec{p}_{i})^2}{2\sum_i m_i} + \sum_{i<j} V_{\text{conf}}(r_{ij}) + \sum_{i<j}
V_\chi(r_{ij}) \, ,
\label{ham}
\end{equation}
with the linear confining interaction :
\begin{equation}
 V_{\text{conf}}(r_{ij}) = -\frac{3}{8}\lambda_{i}^{c}\cdot\lambda_{j}^{c} \, C
\, r_{ij} \, ,
\label{conf}
\end{equation}
and the spin--spin component of the GBE interaction in its $SU_F(3)$ form :
\begin{eqnarray}
V_\chi(r_{ij})
&=&
\left\{\sum_{F=1}^3 V_{\pi}(r_{ij}) \lambda_i^F \lambda_j^F \right.
\nonumber \\
&+& \left. \sum_{F=4}^7 V_{K}(r_{ij}) \lambda_i^F \lambda_j^F
+V_{\eta}(r_{ij}) \lambda_i^8 \lambda_j^8
+V_{\eta^{\prime}}(r_{ij}) \lambda_i^0 \lambda_j^0\right\}
\vec\sigma_i\cdot\vec\sigma_j,
\label{VCHI}
\end{eqnarray}
\noindent
with $\lambda^0 = \sqrt{2/3}~{\bf 1}$, where $\bf 1$ is the $3\times3$ unit
matrix. The interaction (2) contains $\gamma = \pi, K, \eta$ and $\eta '$
meson-exchange terms and the form of $V_{\gamma} \left(r_{ij}\right)$ is given
as the sum of two distinct contributions : a Yukawa-type potential containing
the mass of the exchanged meson and a short-range contribution of opposite
sign, the role of which is crucial in baryon spectroscopy. For a given meson
$\gamma$, the exchange potential is
\begin{equation}V_\gamma (r)=
\frac{g_\gamma^2}{4\pi}\frac{1}{12m_i m_j}
\{\theta(r-r_0)\mu_\gamma^2\frac{e^{-\mu_\gamma r}}{ r}- \frac {4}{\sqrt {\pi}}
\alpha^3 \exp(-\alpha^2(r-r_0)^2)\}
\end{equation}
For the Hamiltonian (2)-(5), we use the
parameters of Refs.\cite{GP96,ST98}. These are :
\begin{eqnarray}
&\frac{g_{\pi q}^2}{4\pi} = \frac{g_{\eta q}^2}{4\pi} =
\frac{g_{Kq}^2}{4\pi}= 0.67,\,\,
\frac{g_{\eta ' q}^2}{4\pi} = 1.206 ,&\nonumber\\
&r_0 = 0.43 \, { fm}, ~\alpha = 2.91 \, { fm}^{-1},~~
 C= 0.474 \, { fm}^{-2}, \, &\nonumber\\& m_{u,d} = 340 \, { MeV}, \, m_s =
440 \, {MeV},&\\
&\mu_{\pi} = 139 \, { MeV},~ \mu_{\eta} = 547 \, { MeV},~
\mu_{\eta'} = 958 \, { MeV},~ \mu_{K} = 495 \, { MeV}.&
\nonumber
\end{eqnarray}
\par

\section{Results}

Values of $\Delta E$ , Eq. (1), for charmed systems
are presented in the table
both for OGE and GBE models. Details of our calculations
based on the GBE model can be found in refs. \cite{SP97,EU98,GE98,SP98}
together with results for $Q = b$.
\begin{table}
\caption[]{Results for $\Delta E$ , Eq.(1), for charmed exotic hadrons}.
\begin{tabular}{|l|l|l|}
\hline
System & OGE & GBE \\
\hline
$u$ $ u$ $\overline{c}$ $\overline{c}$ & 19 MeV \cite{BS93} & -185 MeV \cite
{SP97} \\
 & & \\
$u$ $u$ $d$ $d$ $\overline{c}$ $(P=+1)$ & unbound  &  -76 MeV \cite{EU98}\\
 & & \\
$u$ $u$ $d$ $s$ $\overline{c}$ $(P=-1)$ & -51 MeV \cite{LB89} & 488 Mev
\cite{GE98}\\
 & &\\
$u$ $u$ $d$ $d$ $s$ $c$ & -7.7 MeV \cite{LB93} & 625 MeV \cite{SP98}\\
\hline
\end{tabular}
\end{table}

One can see that the OGE and the GBE interactions predict contradictory
results for the charmed exotic systems presented here: while the GBE
interaction stabilizes a given system, the OGE interaction
destabilizes it and vice versa.  The following remarks are in order:

\begin{itemize}
\item As the $u$ $u$ $d$ $d$ $\overline{c}$ $(P=-1)$ pentaquarks
are predicted to be unbound by a chromomagnetic interaction \cite{LB89},
the same system but with positive parity is expected to be
even more unstable due to the increase in the kinetic energy
produced by the excitation of a quark to the p-shell. While
the OGE interaction favours negative parity pentaquarks
with strangeness,
the best candidates predicted by the GBE interaction
have positive parity and are nonstrange \cite{EU98}.
\item The GBE interaction destabilizes the hexaquarks in the
presence of one or even two heavy quarks \cite{SP98}.
\end{itemize}

\bigskip

\par\vskip 5pt

\acknowledgements
I am grateful to my collaborators
J.-M. Richard, L. Glozman, S. Pepin and M. Genovese who were involved
with me in parts of this work.

\end{document}